\documentclass[twocolumn,aps,preprintnumbers,amsmath,amssymb,superscriptaddress,footinbib]{revtex4}

\usepackage{graphicx,subfigure,multirow}
 \usepackage{longtable}
\usepackage{dcolumn}
\usepackage{bm}
\usepackage[all]{xy}
\usepackage{color}
\usepackage[usenames,dvipsnames]{xcolor}
\usepackage[unicode=true,pdfusetitle,
 bookmarks=true,bookmarksnumbered=false,bookmarksopen=false,citecolor=Turquoise,
 breaklinks=false,pdfborder={0 0 1},backref=false,colorlinks=true,pdfpagemode=FullScreen]
 {hyperref}

\usepackage{ulem}

\makeatletter

\newcommand{\fmslash}[2][0mu]{%
  \mathchoice
    {\fmsl@sh\displaystyle{#1}{#2}}%
    {\fmsl@sh\textstyle{#1}{#2}}%
    {\fmsl@sh\scriptstyle{#1}{#2}}%
    {\fmsl@sh\scriptscriptstyle{#1}{#2}}}
\newcommand{\fmsl@sh}[3]{%
  \m@th\ooalign{$\hfil#1\mkern#2/\hfil$\crcr$#1#3$}}
\makeatother

\newcommand{\lsim}{{\;\raise0.3ex\hbox{$<$\kern-0.75em\raise-1.1ex\hbox{$\sim$}}\;}}
\newcommand{\gsim}{{\;\raise0.3ex\hbox{$>$\kern-0.75em\raise-1.1ex\hbox{$\sim$}}\;}}

\newcommand{\beq}{\begin{equation}}
\newcommand{\eeq}{\end{equation}}
\newcommand{\bea}{\begin{eqnarray}}
\newcommand{\eea}{\end{eqnarray}}
\mathchardef\minus="002D

\usepackage{color}

\begin{document}

\title{Searching for Boosted Dark Matter via Dark-Strahlung}

\author{Doojin Kim}
\email{doojinkim@email.arizona.edu}
\affiliation{Department of Physics, University of Arizona, Tucson, AZ 85721, USA}
\author{Jong-Chul Park}
\email{jcpark@cnu.ac.kr}
\affiliation{Department of Physics, Chungnam National University, Daejeon 34134, Republic of Korea}
\author{Seodong Shin}
\email{seodongshin@yonsei.ac.kr}
\affiliation{Department of Physics \& IPAP, Yonsei University, Seoul 03722, Republic of Korea}

\begin{abstract}
We propose a new search channel for boosted dark matter (BDM) signals coming from the present universe, which are distinct from simple neutrino signals including those coming from the decay or pair-annihilation of dark matter.
The signal process is initiated by the scattering of high-energetic BDM off either an electron or a nucleon.
If the dark matter is dark-sector U(1)-charged, the scattered BDM may radiate a dark gauge boson (called ``dark-strahlung'') which subsequently decays to a Standard Model fermion pair.
We point out that the existence of this channel may allow for the interpretation that the associated signal stems from BDM, not from the dark-matter-origin neutrinos.
Although the dark-strahlung process is generally subleading compared to the lowest-order simple elastic scattering of BDM, we find that the BDM with a significant boost factor may induce an $\mathcal{O}(10-20\%)$ event rate in the parameter regions unreachable by typical beam-produced dark-matter.
We further find that the dark-strahlung channel may even outperform the leading-order channel in the search for BDM, especially when the latter is plagued by substantial background contamination.
We argue that cosmogenic BDM searches readily fall in such a case, hence taking full advantage of dark-strahlung.
As a practical application, experimental sensitivities expected in the leading-order and dark-strahlung channels are contrasted in dark gauge boson parameter space, under the environment of DUNE far-detectors, revealing usefulness of dark-strahlung.
\end{abstract}

\maketitle

\section{Introduction}

The models of boosted dark matter (BDM) are receiving rising attention~\cite{Kachulis:2017nci, Ha:2018obm} as an alternative scenario to reconcile the paradigm of thermal dark-matter with the null observation of dark-matter-induced signatures via non-gravitational interactions~\cite{Belanger:2011ww, Agashe:2014yua}.
Many of them predict that some (subdominant) dark-matter components can be substantially boosted in the present universe and manifest themselves by relativistic scattering signatures in terrestrial detectors.
The most minimal experimental signature arising in this class of models is the recoil of either an electron or a nucleon induced by the elastic scattering of BDM, so a number of preceding works have focused on the simple possibility in various experiments, assuming diverse dark-sector scenarios such as scalar/fermionic dark-matter with scalar/vector portals~\cite{Agashe:2014yua, Huang:2013xfa, Bhattacharya:2014yha, Berger:2014sqa, Kong:2014mia,Cherry:2015oca, Kopp:2015bfa, Necib:2016aez, Alhazmi:2016qcs, Kim:2018veo,Berger:2018urf}.
By contrast, Refs.~\cite{Kim:2016zjx, Giudice:2017zke, Chatterjee:2018mej, Aoki:2018gjf, Heurtier:2019rkz} explored the potential of cosmogenic BDM-induced inelastic upscattering, adding unstable particles to the underlying dark sector as was done in inelastic dark-matter models~\cite{TuckerSmith:2001hy}.

The latter possibility is advantageous in the search for BDM signals~\cite{Kim:2016zjx, Giudice:2017zke, Chatterjee:2018mej}, as they usually accompany additional visible particles together with an electron/nucleon recoil.\footnote{A similar point was made in Ref.~\cite{Izaguirre:2014dua} in the context of fixed target experiments.}
While the elastic BDM signature searches often suffer from large background contamination (e.g., atmospheric neutrinos), the inelastic ones may enjoy even (nearly) zero-background environment due to many signal features, at the {\it expense} of ``minimalism'' of underlying BDM models.

Indeed, it is noteworthy that such distinctive signatures may arise even under the minimal setup, once higher-order corrections are taken into account. 
As an example of vector-portal type scenarios, a dark gauge boson can radiate from Standard Model (SM) fermions and/or dark-sector matter particles just like the ordinary bremsstrahlung.
Ref.~\cite{Bjorken:2009mm} studied the so-called ``$A'$-strahlung'' of electron in beam-dump experiments, which is suppressed by the kinetic mixing.
Very recently, Ref.~\cite{deGouvea:2018cfv} proposed the process of ``dark trident'' in which beam-produced dark-matter scatters off a target nucleus, emitting a dark gauge boson which subsequently disintegrates to a SM fermion pair.
The resultant experimental signature is similar to that of the ordinary neutrino trident, allowing for almost background-free searches.
Related phenomenology was also studied at the LHC~\cite{Gupta:2015lfa, Bai:2015nfa, Zhang:2016sll}.

\begin{figure*}[t]
\centering
\includegraphics[width=18cm]{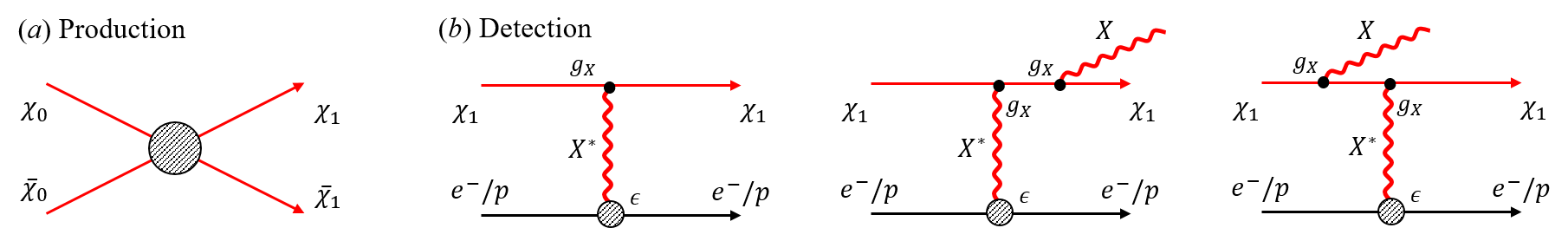}
\caption{\label{fig:diagram} The processes under consideration: ($a$) production of BDM in the present universe, ($b$) detection of BDM via leading-order scattering, final state dark-strahlung (FSDS), and initial state dark-strahlung (ISDS) from left to right.
}
\end{figure*}

Inspired by the above approaches investigated in beam-dump experiments, in this paper, we propose a {\it new} BDM search strategy utilizing dark gauge boson bremsstrahlung from cosmogenic BDM.
As delineated in FIG.~\ref{fig:diagram}($b$), a dark gauge boson can be emitted by either the scattered or incident BDM.
We separately label them by final state dark-strahlung (FSDS) and initial state dark-strahlung (ISDS).
In principle, the incoming and outgoing target particles can radiate a dark gauge boson, i.e., ``$A'$-strahlung'', but the contribution should be at least subleading unless the dark-sector gauge coupling is as small as the kinetic mixing.

Remarkably, the existence of this dark-strahlung channel carries a profound physics implication.
BDM usually behaves like neutrinos, resulting in the experimental signatures which would be invoked by neutrinos.
Therefore, a discovery of the BDM-like signal would create a challenging task to verify that observed events are actually BDM-induced, rejecting the hypothesis of neutrinos from the decay/pair-annihilation of cosmological dark-matter.
Since neutrinos do {\it not} involve this sort of process (except $W/Z$-strahlung although negligible), the additional observations of BDM signals in this channel might serve as a clue to cosmic-origin, energetic dark-matter models.

Despite the distinctive signal features, however, the dark-strahlung process bears an innate drawback that its production cross section is smaller than the corresponding leading-order (LO) contribution [see the first diagram in FIG.~\ref{fig:diagram}($b$)].
The relative magnitude depends on the dark-sector coupling constant,
the incoming energy and mass of BDM, and the mass of dark gauge boson.
We shall first perform a quantitative examination of the dark-strahlung cross sections relative to the leading-order ones in the variation of these model parameters, and show that the dark-strahlung-induced contribution can reach $\mathcal{O}(10\%)$ of the leading order for the BDM carrying an energy of a few tens of GeV.
We emphasize that such energetic dark-matter readily arises in various BDM scenarios, whereas the typical energy scale of beam-produced dark-matter in existing beam facilities is limited to $\lesssim \mathcal{O}(10\hbox{ GeV})$.

More importantly, the usefulness of dark-strahlung can be further highlighted, especially when the leading-order-associated signal channel is overwhelmed by backgrounds and/or such backgrounds are not well under control.
In fact, the cosmogenic BDM searches fall in this case so that they often benefit significantly from dark-strahlung.
As a measure to quantify this statement, we shall compare the required amount of data to achieve a given level of experimental sensitivity between the leading-order and the dark-strahlung channels, showing that the dark-strahlung channel is comparable, competitive, or even superior to the leading-order one, in a wide range of BDM parameter space.

We further study the impact of realistic effects such as cuts and acceptance of the events.\footnote{The details are described in Ref.~\cite{Giudice:2017zke}.}
To this end, we shall employ the far-detector environment in Deep Underground Neutrino Experiment (DUNE)~\cite{Abi:2018dnh,Abi:2018alz,Abi:2018rgm}, follow the analysis strategy in Ref.~\cite{Chatterjee:2018mej}, and demonstrate that the dark-strahlung channel may outperform the leading-order one in probing certain regions of dark gauge boson parameter space.

The paper is organized as follows.
In Sec.~\ref{sec:model} we briefly define our benchmark BDM scenario, followed by the discussion on the signal and background processes.
Sec.~\ref{sec:pheno} is devoted to our major findings including the comparisons of production cross sections and signal statistics required for 90\% confidence level (C.L.) experimental sensitivity between the dark-strahlung and the leading-order, and the experimental reach in dark gauge boson parameter space which would be achieved in the DUNE.
Our conclusions and future prospects are summarized in Sec.~\ref{sec:conclusion}.

\section{Model Setup \label{sec:model}}

Various scenarios to create relativistically moving dark-matter in the present universe have been suggested~\cite{Agashe:2014yua, Kim:2016zjx, DEramo:2010keq, Belanger:2011ww, Huang:2013xfa, Bhattacharya:2014yha, Berger:2014sqa, Kong:2014mia,Cherry:2015oca, Kopp:2015bfa, Necib:2016aez, Alhazmi:2016qcs, Kim:2017qaw, Giudice:2017zke, Chatterjee:2018mej, Kim:2018veo, Berger:2018urf, Aoki:2018gjf, Heurtier:2019rkz, Yin:2018yjn, Bringmann:2018cvk, Ema:2018bih}.
Of them, we choose the two-component dark-matter model proposed in~\cite{Belanger:2011ww, Agashe:2014yua} as our benchmark scenario, in which a pair of the heavier components $\chi_0$ of mass $m_0$ annihilate to a pair of the lighter components $\chi_1$ of mass $m_1$, giving a Lorentz boost factor for $\chi_1$ by the mass ratio $m_0/m_1$ [see also FIG.~\ref{fig:diagram}($a$)].
An example operator to account for such an annihilation is given by
\bea
\mathcal{L}_{\rm int} \supset \frac{1}{\Lambda^2} \bar{\chi}_0\chi_0 \bar{\chi}_1\chi_1\,, \label{eq:lag1}
\eea
where $\Lambda$ encodes relevant high-scale physics whose details are irrelevant in this study.
Just for concreteness, we assume that the two dark-matter species are Dirac-fermionic.
The expected flux of BDM $\chi_1$ from the galactic halo can be estimated as~\cite{Agashe:2014yua}
\bea
\mathcal F(\theta_C) \hspace{-0.2cm} &=& \hspace{-0.2cm} \frac{1}{2}\cdot \frac{1}{4\pi}\int_{\theta < \theta_C} d\Omega \int_{\rm los}ds \langle \sigma v \rangle_{0\to 1}\left\{\frac{\rho(s,\theta)}{m_0} \right\}^2 \nonumber \\
\hspace{-0.2cm} & \stackrel{\theta_C \to 180^{\circ}}{\longrightarrow} & \hspace{-0.2cm}1.6 \times 10^{-4}\,{\rm cm}^{-2} {\rm s}^{-1} \nonumber \\
\hspace{-0.2cm} && \hspace{-0.2cm} \times \, \left( \frac{\langle \sigma v \rangle_{0 \to 1}}{5 \times 10^{-26}\,{\rm cm}^3\, {\rm s}^{-1}} \right) \times \left( \frac{{\rm GeV}}{m_0}\, \right)^2, \label{eq:fluxformula}
\eea
where $\rho$ is the $\chi_0$ density distribution in terms of the line-of-sight (los) $s$ and $\langle \sigma v \rangle_{0 \to 1}$ is the thermal-averaged annihilation cross section of $\chi_0 \overline{\chi}_0 \to \chi_1 \overline{\chi}_1$ at the universe today.
$\theta_C$ denotes a cone angle around the direction of the galactic center, so restricting the polar angle to $\theta<\theta_C$ allows for finding the flux within the $\theta_C$ cone.
The reference flux in the second line results from $\theta_C=180^{\circ}$ and the Navarro-Frenk-White dark-matter halo profile~\cite{Navarro:1995iw, Navarro:1996gj} with local dark-matter density near the Sun being $0.3\,{\rm GeV}{\rm cm}^{-3}$.

To describe interactions between $\chi_1$ and SM particles, we adopt a simplified model based on the dark gauge boson mixing scenario~\cite{Okun:1982xi, Galison:1983pa, Holdom:1985ag, Chun:2010ve, PARK:2016wip}:
\bea
-\mathcal{L}_{\rm int} \supset \frac{\epsilon}{2}F_{\mu\nu}X^{\mu\nu} + g_X \bar{\chi}_1 \gamma^{\mu}\chi_1 X_\mu\,,~~ \label{eq:lag2}
\eea
where $\epsilon$ and $g_X$ denote the kinetic mixing parameter and the dark-sector coupling constant, respectively.
$F^{\mu\nu}$ and $X^{\mu\nu}$ are the field strength tensors for the SM photon and the dark gauge boson.
The second term in~\eqref{eq:lag2} allows for dark-strahlung, and the emitted dark gauge boson predominantly decays to a pair of SM fermions through the first term in~\eqref{eq:lag2} if the following mass hierarchy holds:
\bea
m_X <2 m_1\,.
\eea
If the incoming energy of BDM $\chi_1$ (defined by $E_1$) is sufficiently large, the center-of-mass energy $\sqrt{s}$ is $\sim \sqrt{2E_1m_T}$ with $m_T$ being the mass of target particle.
For electron recoil-involving events with $E_1\sim 100$ GeV, $\sqrt{s}\sim300$ MeV, so the di-muonic final state is kinematically accessible only if the radiating $X$ is {\it off}-shell.\footnote{Obviously, this is not the case for the proton target, as it allows for a much bigger $\sqrt{s}$ than the electron target for a given $E_1$.}
However, such a possibility is highly suppressed, hence nearly negligible compared to the leading-order contribution.
Simply for illustration, we consider only the electron target hereafter, while our argument is straightforwardly applicable to the proton target.
Therefore, a dark-strahlung-induced event consists of a recoil electron and an electron-positron pair in the final state.

Speaking of the backgrounds, the dominant source against the leading-order scattering process is the atmospheric neutrino.
Assuming a uniform flux from the entire sky, we see that the number of atmospheric $\nu$-induced background events within a cone of angle $\theta_C$ is given by
\bea
N_{\rm BG}(\theta_C) = \frac{1-\cos\theta_C}{2} \times N_{\rm BG}(180^\circ)\,. \label{eq:Nallsky}
\eea
For DUNE-like detectors with recoil electron energy greater than 30 MeV, we estimate that the total number over the entire sky is~\cite{Acciarri:2015uup}
\bea
N_{\rm BG}(180^\circ) \simeq 40.2\, {\rm yr}^{-1} {\rm kt}^{-1}\,. \label{eq:Ntheta}
\eea

On the other hand, it is expected that the dark-strahlung-induced signal event hardly suffers from background contamination due to several distinctive features such as multiple visible particles and (potentially) delayed decay of radiating dark gauge bosons.
In particular, detectors with good particle identifications and resolutions \footnote{A representative one is the liquid argon time projection chamber (LArTPC)-based detector which is generally characterized by good particle identifications and good angular, energy, vertex, and time resolutions.} may fully benefit from such advantages~\cite{Izaguirre:2014dua, Kim:2016zjx, Giudice:2017zke, Chatterjee:2018mej}.
Indeed, one may imagine several possibilities which would appear signal-like.
A highly plausible situation is that an atmospheric neutrino provokes a resonance or deep inelastic scattering process, resulting in a handful of mesons (usually pions) whose visible decay products may collectively leave a signal-like signature.
To carefully estimate such events, we combine the electron-neutrino flux studied by the Super-Kamiokande Collaboration~\cite{Honda:2015fha} and the neutrino cross sections from Ref.~\cite{Formaggio:2013kya}, and find that $\sim 12$ events per kt$\cdot$yr will be relevant.\footnote{Similar considerations were made for the dark-trident process in Ref.~\cite{deGouvea:2018cfv} in terms of neutrino trident~\cite{Altmannshofer:2014pba}, charged-current $\nu_e$, neutral-current $\nu_\mu$, etc.}
However, we expect that detectors with good particle identification, e.g., LArTPC~\cite{Acciarri:2016sli,Acciarri:2017hat,Acciarri:2018ahy}, will separate out such fake events very efficiently.
In addition, the techniques of machine-learning-based particle identification have been developed in the field, e.g., Refs.~\cite{Acciarri:2016ryt,Adams:2018bvi}.
Given all these deliberations and careful estimates, we anticipate that the dark-strahlung channel will be nearly background-free at quality-performance detectors so that relevant searches can be conducted under a zero-background assumption.

\section{Phenomenology \label{sec:pheno}}

In this section, we discuss various phenomenological implications of dark-strahlung.
We begin with the comparison between leading order production cross section and dark-strahlung processes, followed by a discussion of dark-strahlung for beam-produced and cosmogenic dark-matter.
The last subsection demonstrates an application of the dark-strahlung channel, using the DUNE far-detectors as an example.


\subsection{Simple elastic scattering vs. dark-strahlung}

\begin{figure*}[t]
\centering
\includegraphics[width=8.0cm]{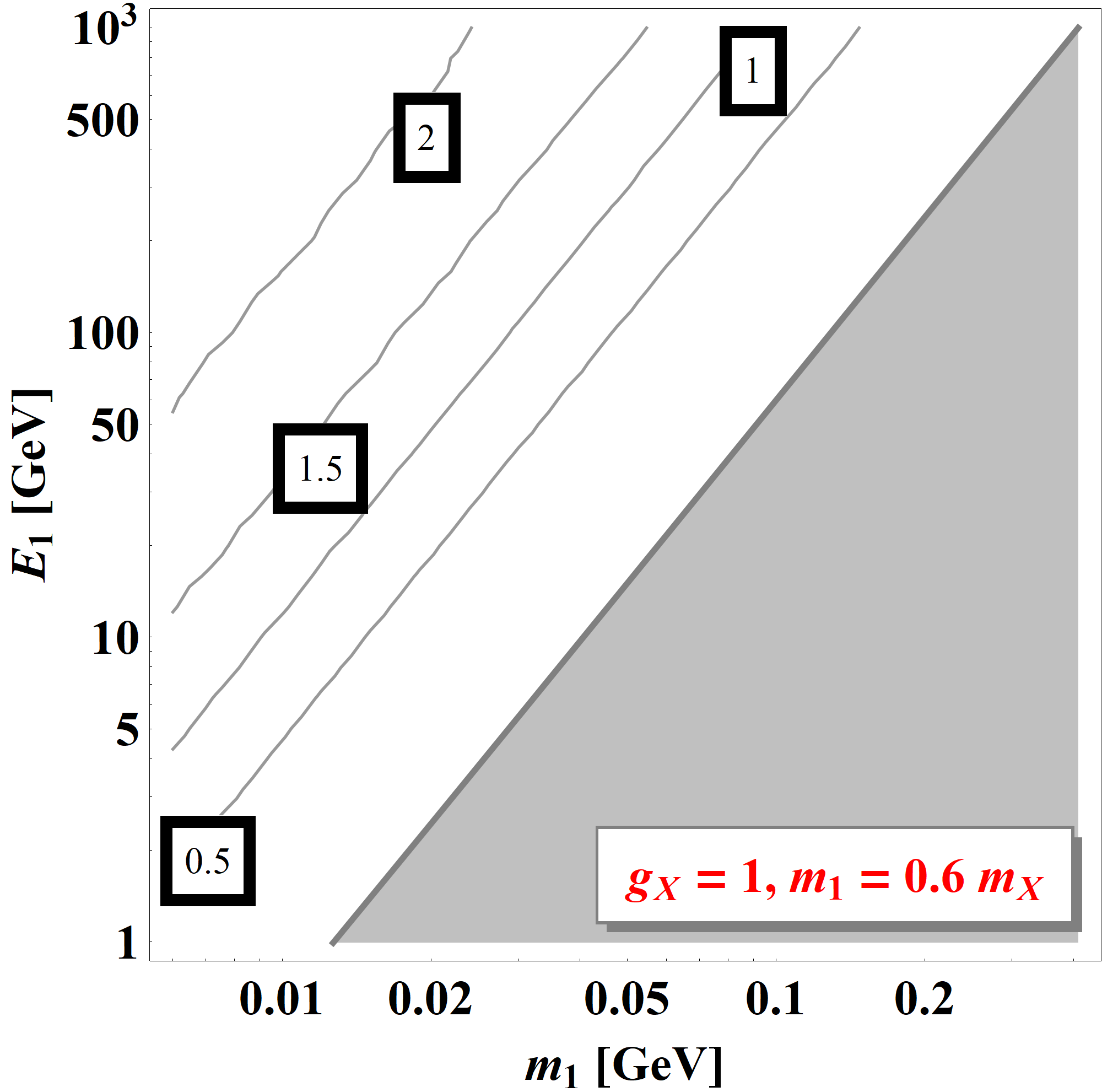}\hspace{0.5cm}
\includegraphics[width=8.0cm]{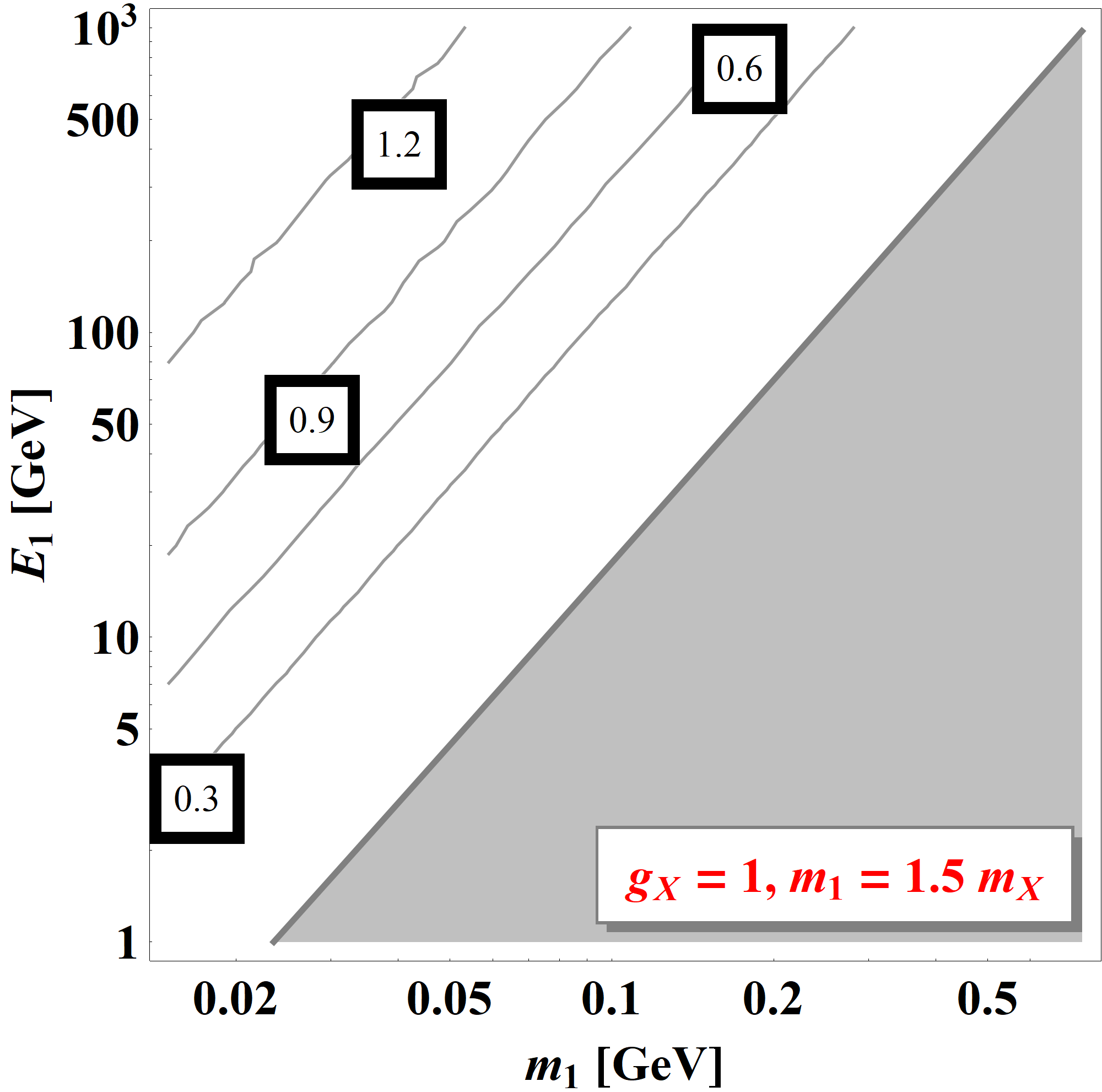}

\vspace{0.5cm}

\includegraphics[width=8.0cm]{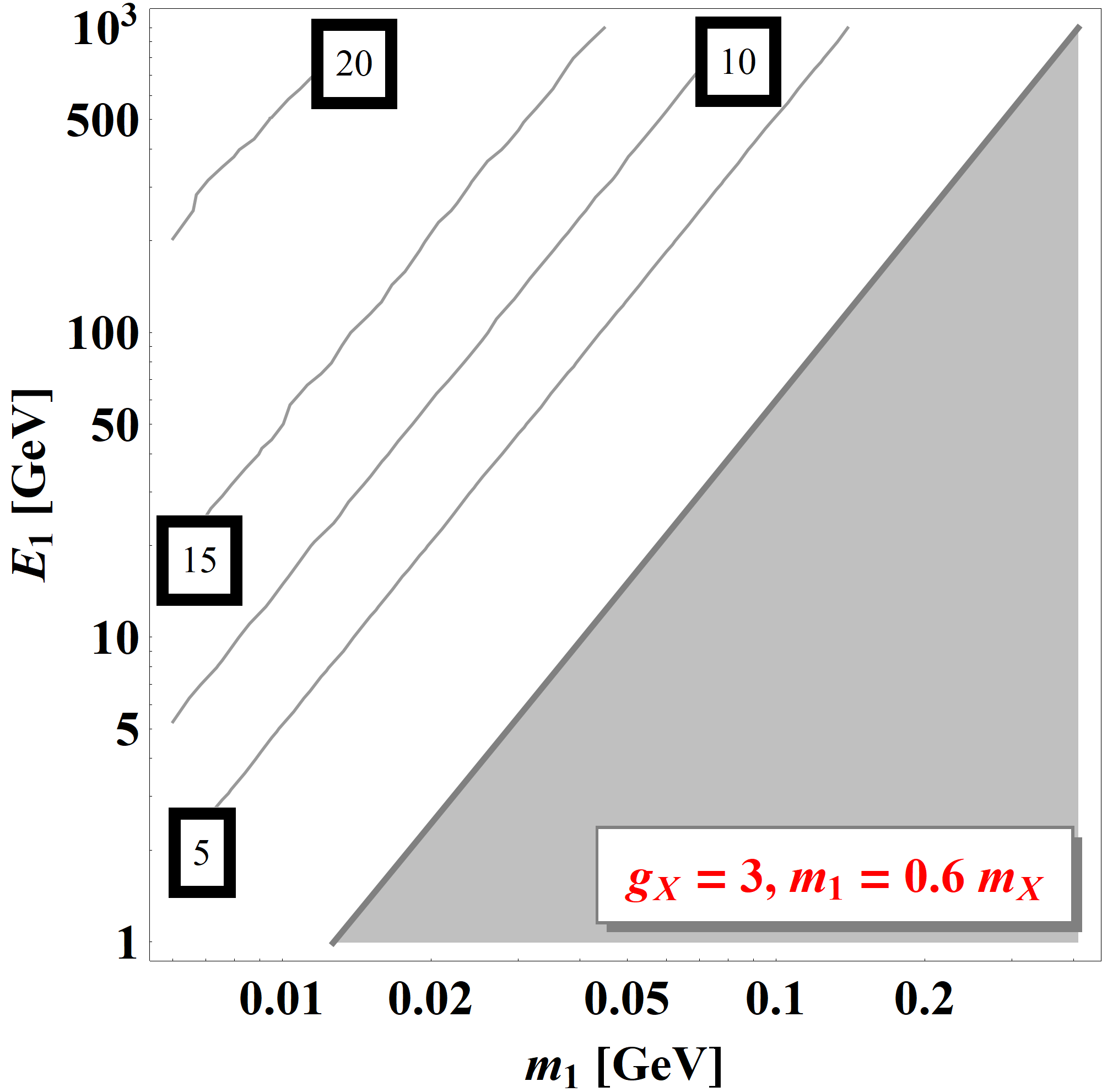}\hspace{0.5cm}
\includegraphics[width=8.0cm]{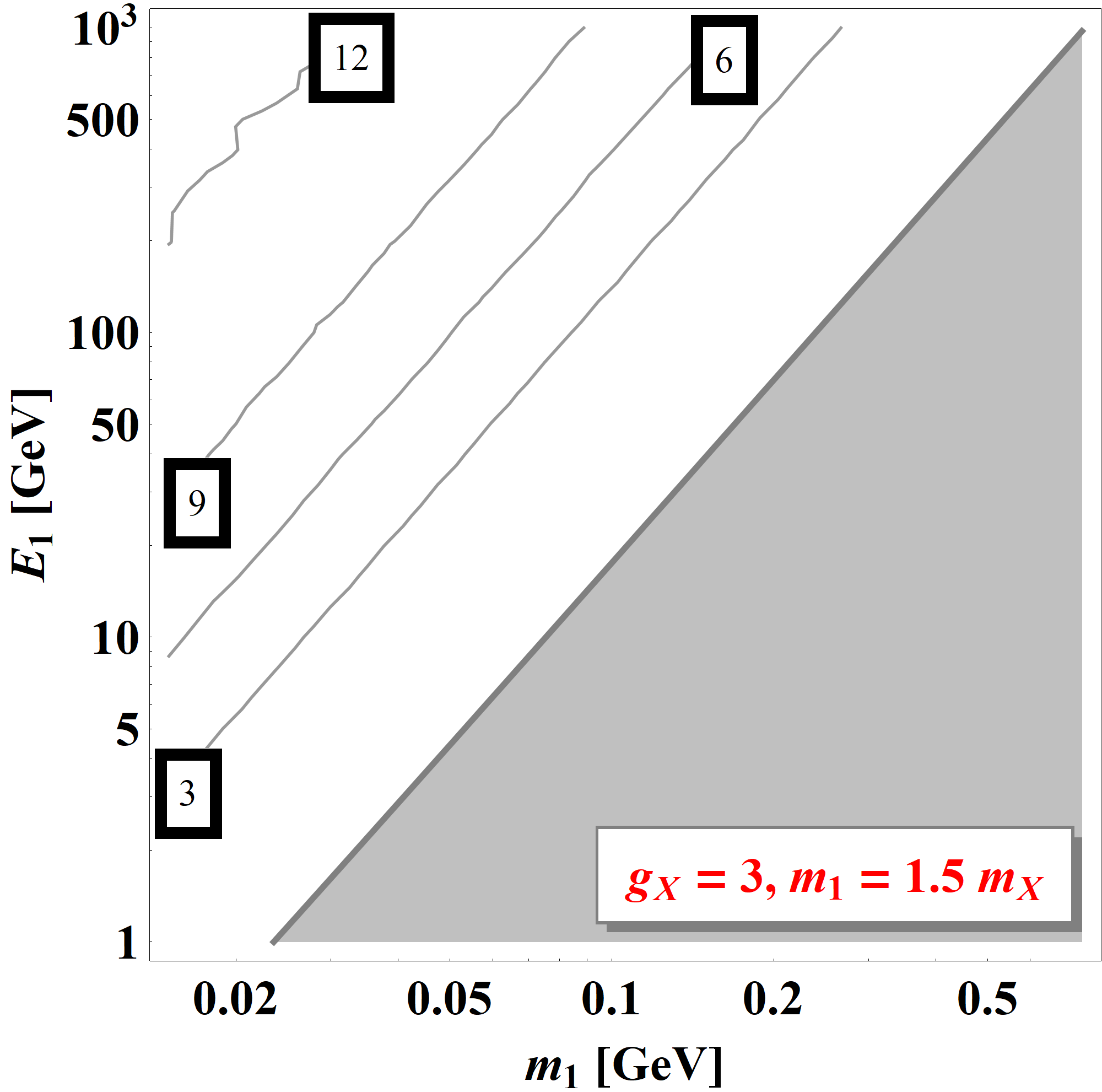}
\caption{\label{fig:XSRatio}
Cross-section ratios between the leading-order and the dark-strahlung processes, $\sigma_{\rm DS}/\sigma_{\rm LO}$, in the $m_1 - E_1$ plane.
The top (bottom) panels are for $g_X=1$ ($g_X=3$), while the left (right) panels are for $m_1/m_X=0.6$ ($m_1/m_X=1.5$).
Contours correspond to the calculated cross-section ratios in percent [\%].
The gray-shaded regions show the kinematically inaccessible parameter space for the {\it on}-shell dark-strahlung process.
Note that the $\epsilon$ choice is not specified as the dependence is exactly canceled out.
The dark-strahlung events occur more frequently in decreasing $m_1$, $m_X$ and increasing $g_X$, $E_1$.
}
\end{figure*}

We first compare the production cross sections of the dark-strahlung process, $\sigma_{\rm DS}$, relative to those of the leading-order process, $\sigma_{\rm LO}$, with variations of model parameters.
The cross sections are numerically calculated with \texttt{MG5\_aMC@NLO}~\cite{Alwall:2014hca}.
FIG.~\ref{fig:XSRatio} displays cross-section ratios, $\sigma_{\rm DS}/\sigma_{\rm LO}$, in the $m_1-E_1$ plane.
The top (bottom) panels are for $g_X=1$ ($g_X=3$), while the left (right) panels are for $m_1/m_X=0.6$ ($m_1/m_X=1.5$).
Contours show the cross-section ratios in percent [\%].
The gray-shaded regions correspond to the kinematically inaccessible parameter space for the {\it on}-shell dark-strahlung process.
Note that we do not specify $\epsilon$ choices as the dependence is exactly canceled out.

The results suggest that the dark-strahlung events occur more frequently in decreasing $m_1$, $m_X$ and increasing $g_X$, $E_1$ as expected in QED.
Very roughly, the dark-strahlung contribution relative to the leading-order is suppressed by $\sim \alpha_X/\pi$ ($4\pi \alpha_X \equiv g_X^2$), while detailed ratios depend on model parameters.
We find from this numerical study that $\sigma_{\rm DS}$ could be even $\mathcal{O}(10-20\%)$ of $\sigma_{\rm LO}$ with $E_1\sim 0.1 - 1$ TeV if the underlying dark-sector gauge coupling is as sizable as $\mathcal{O}(1)$ (see e.g., Ref.~\cite{Tulin:2017ara} and references therein for dark-sector scenarios with large couplings).

\begin{figure*}[t]
\centering
\includegraphics[width=8.0cm]{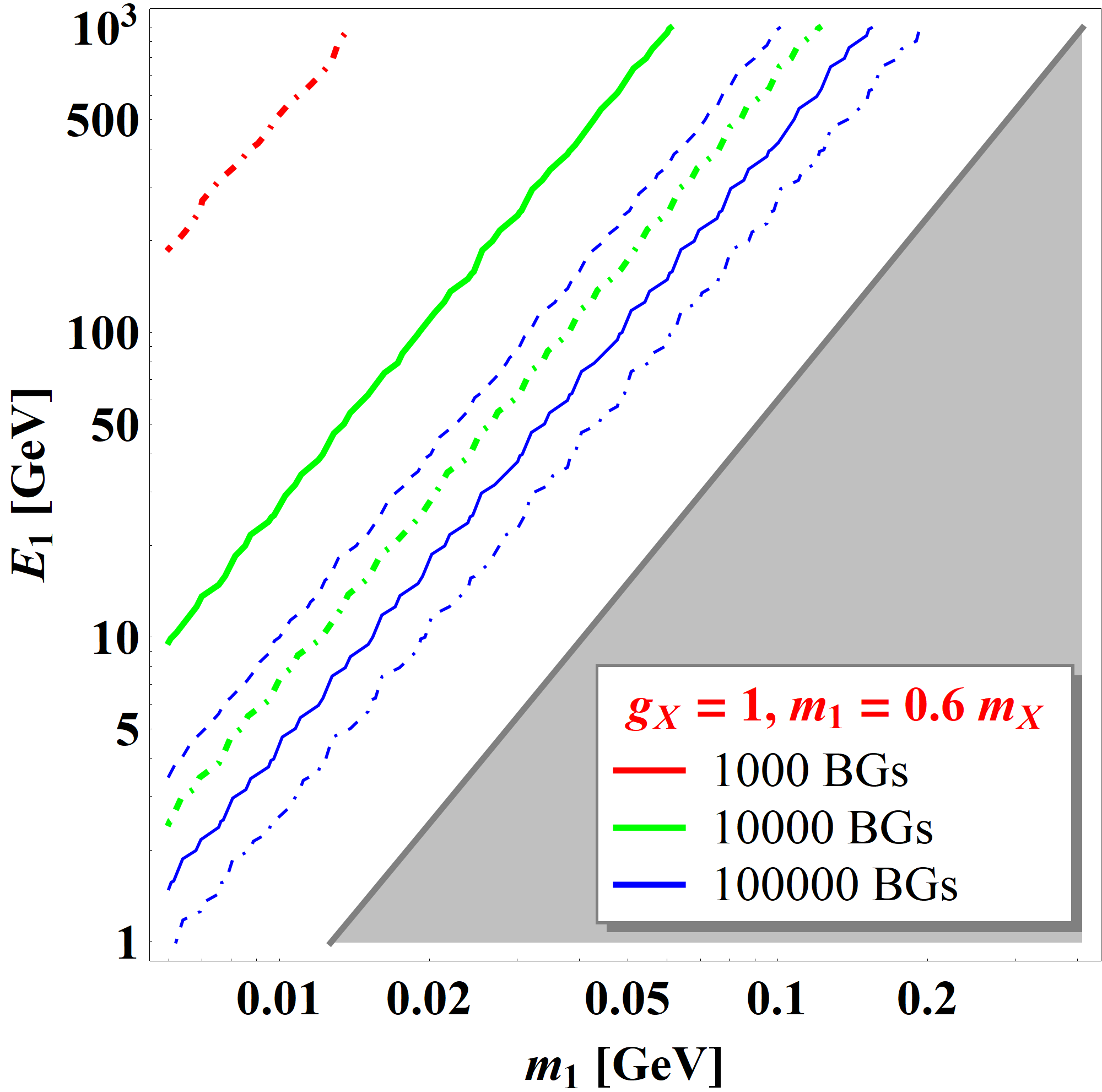}\hspace{0.5cm}
\includegraphics[width=8.0cm]{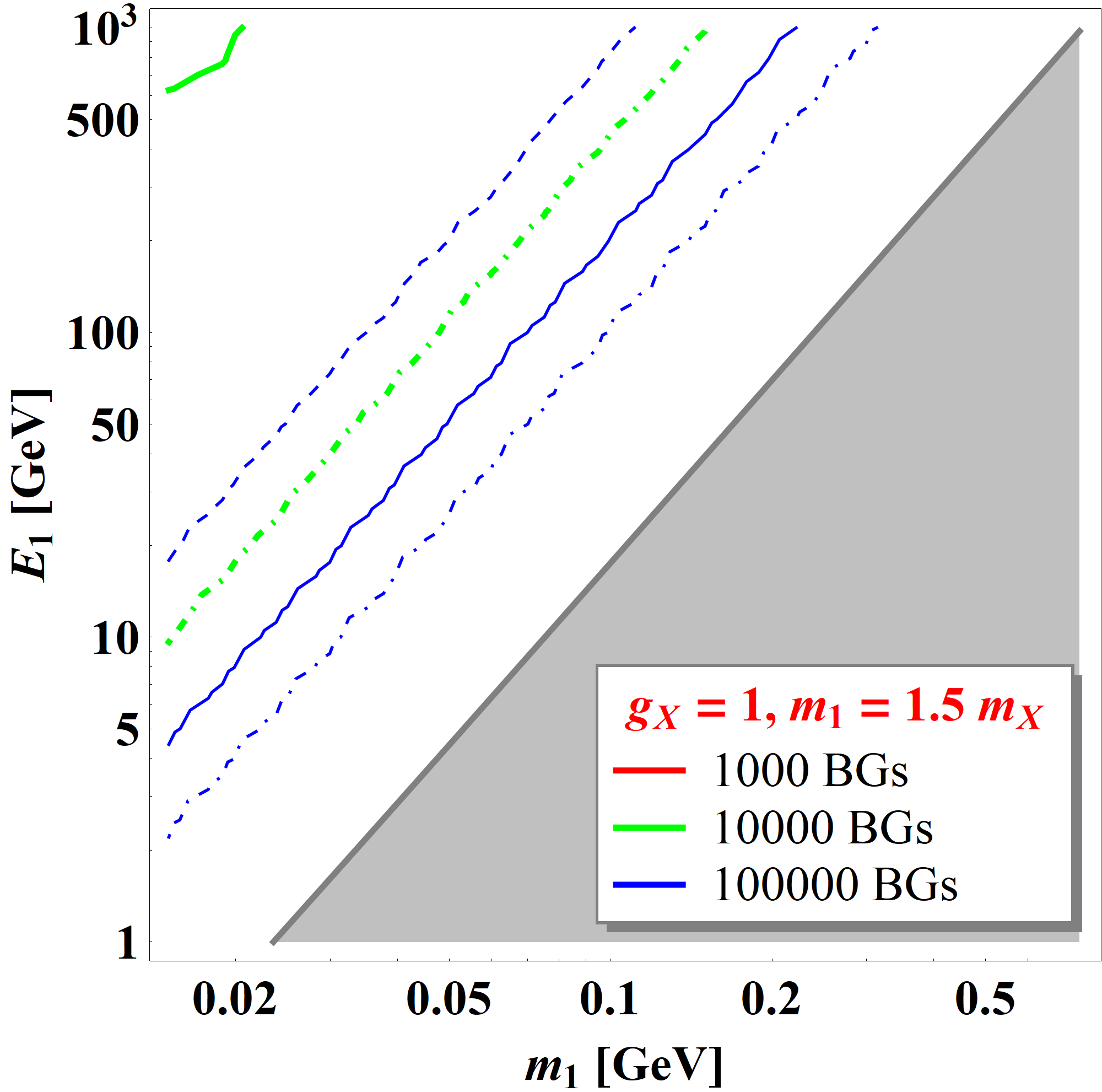}

\vspace{0.5cm}

\includegraphics[width=8.0cm]{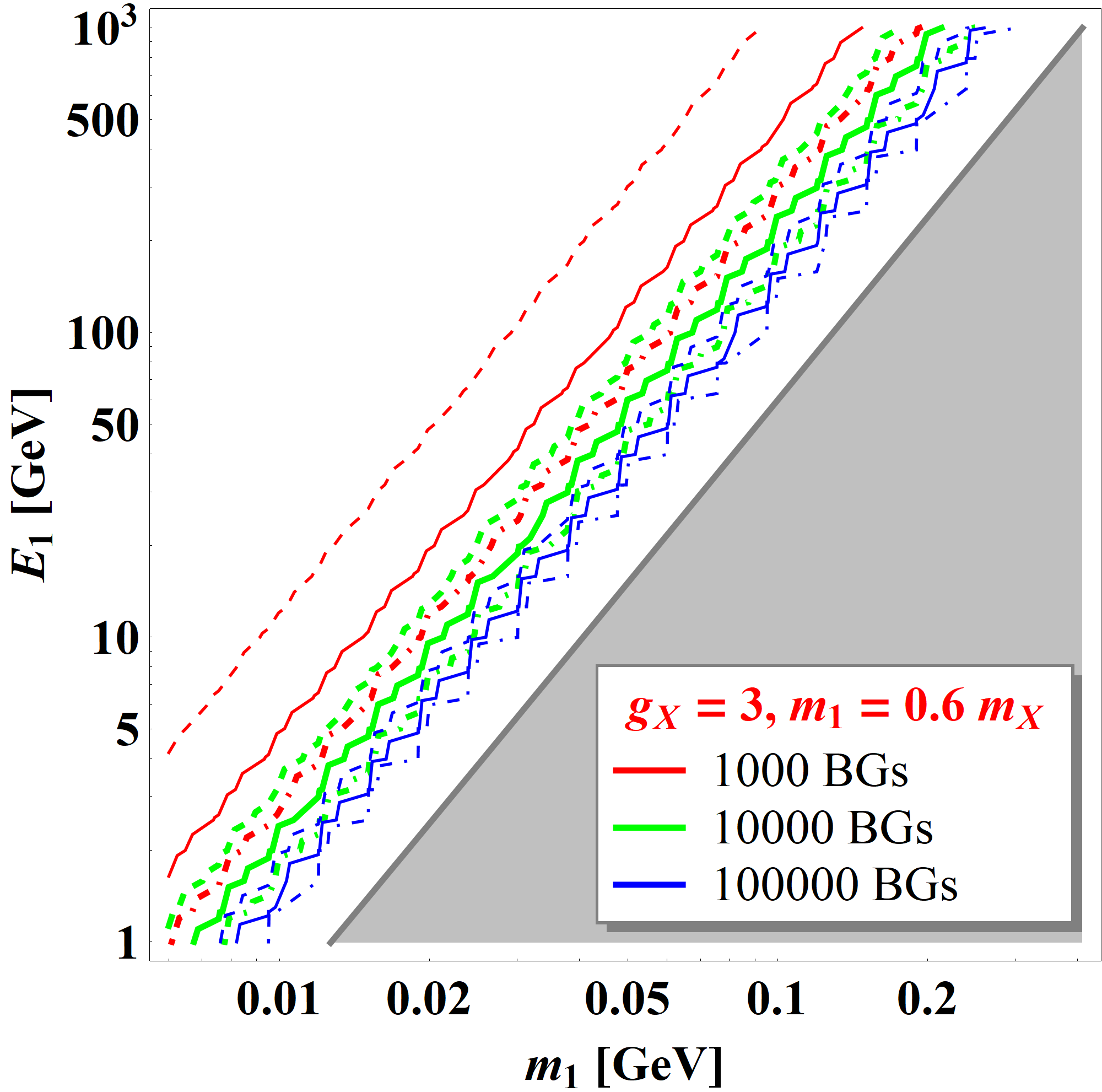}\hspace{0.5cm}
\includegraphics[width=8.0cm]{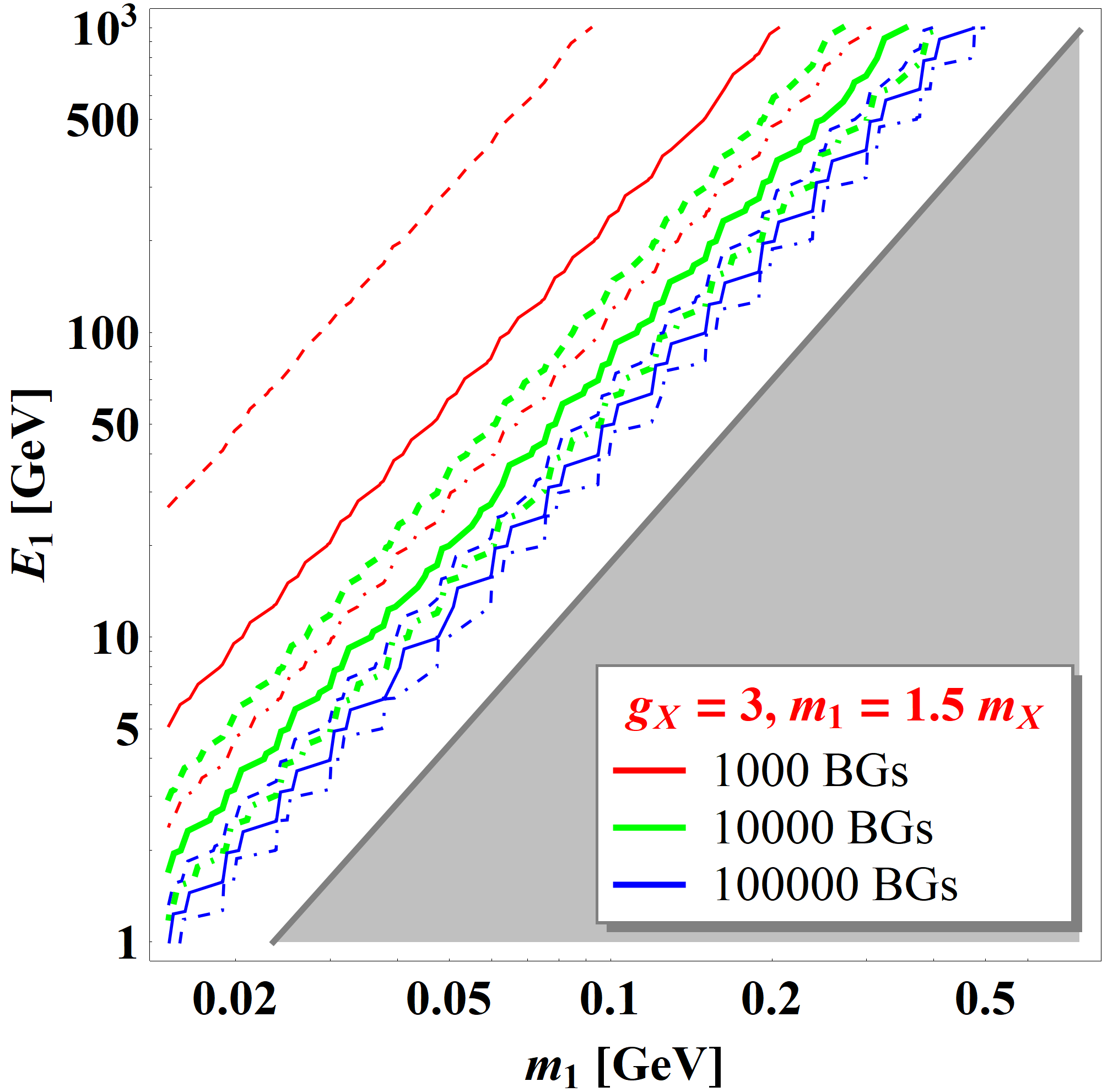}
\caption{\label{fig:fluxRatio}
Ratios of required run-times to achieve 90\% C.L., $T_{\rm DS}/T_{\rm LO}$, in the $m_1 - E_1$ plane.
Three reference numbers of backgrounds are assumed for the leading-order process: $N_{\rm BG} = 10^3, 10^4$, and $10^5$ for red, green, and blue curves, respectively.
The parameter choices are the same as in FIG.~\ref{fig:XSRatio}.
Dashed, solid, and dot-dashed curves denote $T_{\rm DS}/T_{\rm LO}=0.5, 1,$ and 2, correspondingly.
}
\end{figure*}

However, the true potential of dark-strahlung cannot be assessed without involving backgrounds since typical BDM searches via the simple leading-order process encounter the issue of enormous signal-mimicking events.
As a measure, we introduce the ratio of the required run-time to achieve the (statistical-uncertainty-based) 90\% C.L. sensitivity for a given model point, $T_{\rm DS}/T_{\rm LO}$.\footnote{
More precisely, the measure comes from the numbers of required signal events to achieve the 90\% C.L. sensitivity, $N_{\rm DS}^{90}$ and $N_{\rm LO}^{90}$.
$N^{90}$ varies by the associated background assumption.
The number of events is given by $\sigma_{\rm DS/LO}\cdot \mathcal{F}\cdot A\cdot N_e \cdot T_{\rm DS/LO}$, where $\mathcal{F}$ and $N_e$ are the flux of $\chi_1$ and the number of target electrons in a given detector.
$A$, a probabilistic quantity, encodes all experimental details such as efficiency and acceptance.
We here assume an ideal detector resulting in $A=1$.
Therefore, we have \\
$$\frac{T_{\rm DS}}{T_{\rm LO}} = \frac{N_{\rm DS}^{90}/(\sigma_{\rm DS} \mathcal{F}AN_e)}{N_{\rm LO}^{90}/(\sigma_{\rm LO} \mathcal{F}AN_e)}=\frac{N_{\rm DS}^{90}/\sigma_{\rm DS}}{N_{\rm LO}^{90}/\sigma_{\rm LO}}\,.$$
\\
See Refs.~\cite{Kim:2018veo, Giudice:2017zke} for more related details.
}
For example, $T_{\rm DS}/T_{\rm LO}=2$ means that the dark-strahlung channel requires twice more run-time than the leading-order does.
FIG.~\ref{fig:fluxRatio} reports our results for $T_{\rm DS}/T_{\rm LO}$ in the $m_1 - E_1$ plane.
Three reference background assumptions are taken for the leading-order process, $N_{\rm BG} = 10^3$ (red), $10^4$ (green), and $10^5$ (blue), while zero backgrounds are assumed for the dark-strahlung one.
The parameter choices are identical to those in FIG.~\ref{fig:XSRatio}.
A few contours are included as a guidance, $T_{\rm DS}/T_{\rm LO}=0.5$, 1, and 2 by dashed, solid, and dot-dashed curves, respectively.

The results can be understood as follows. 
First of all, we see that even under mild background contamination (i.e., $N_{\rm BG}=10^3$) in the leading-order channel, the dark-strahlung channel remains competitive in a wide range of parameter space as it requires a similar amount of run-time.
Similar expectations can be extended to the discovery potentials.
For example, imagine a scenario that underlying physics comes with $g_X=1$ and highly-boosted $\chi_1$ of $\mathcal{O}(10 \hbox{ MeV})$ mass and the leading-order channel is tied to $\mathcal{O}(10^3)$ background events (see also the top-left panel of FIG.~\ref{fig:fluxRatio}).
In this case, if a sufficient number of BDM-looking events were recorded within one-year time exposure, then one needs to wait for only a few years to confirm or refute the BDM hypothesis!

\subsection{Beam-produced dark-matter vs. cosmogenic BDM}

It is informative to discuss and contrast implications of dark-strahlung (or similarly ``dark trident'') for beam-produced dark-matter and cosmogenic BDM at this point, as both deal with relativistic dark-matter.
The beam energy is distributed to all produced particles including dark matter, and thus individual beam-produced dark-matter typically carries away an energy much less than the original beam energy, for example, $\sim 10$ GeV ($\sim20$ GeV) for a 120 (400) GeV beam~\cite{deNiverville:2016rqh}.
By contrast, many mechanisms for producing BDM in the present universe allow $E_1$ to be $\mathcal{O}(100\hbox{ GeV})$ or even greater than 1 TeV without losing much signal statistics~\cite{Berger:2014sqa, Kong:2014mia, Yin:2018yjn}.
Our results in FIG.~\ref{fig:XSRatio} unambiguously suggest that the dark-strahlung contributions can be above $\mathcal{O}(10\%)$ in such high-energy realm, and therefore, the search for cosmic-origin signals enjoy (presumably) {\it unexpected} effects from dark-strahlung.

Moreover, the results in FIG.~\ref{fig:fluxRatio} imply that the more the leading-order channel suffers from backgrounds, the more considerable the dark-strahlung channel is.
In general, the backgrounds to the leading-order process (e.g. atmospheric-neutrino-induced, single, $e$-like events for this case) are not well under control in the cosmic-origin BDM searches, in comparison with the beam-produced dark-matter searches in particle accelerator experiments which are more capable of controlling potential backgrounds, e.g., on-/off-target beam data analyses in MiniBooNE~\cite{Aguilar-Arevalo:2017mqx, Aguilar-Arevalo:2018wea}.
This actually has an implication for surface-based detectors, e.g., ProtoDUNE and SBN, which additionally suffer from overwhelming cosmic-induced backgrounds.
While ideas of inelastic BDM searches~\cite{Chatterjee:2018mej} or ``Earth Shielding''~\cite{Kim:2018veo} allow to get around the background issue, we expect that the dark-strahlung channel will provide a new chance in the search for minimal BDM signals in the surface-based experiments.

\subsection{Experimental sensitivities of DUNE}

\begin{figure*}[t]
\centering
\includegraphics[width=8.0cm]{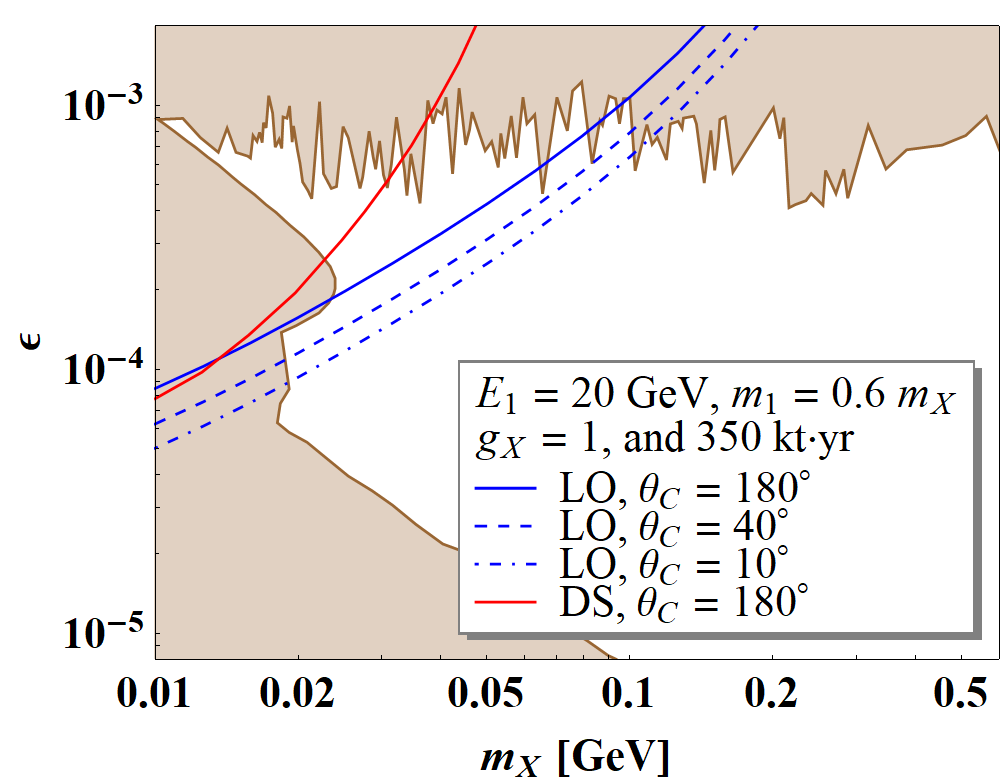}\hspace{0.5cm}
\includegraphics[width=8.0cm]{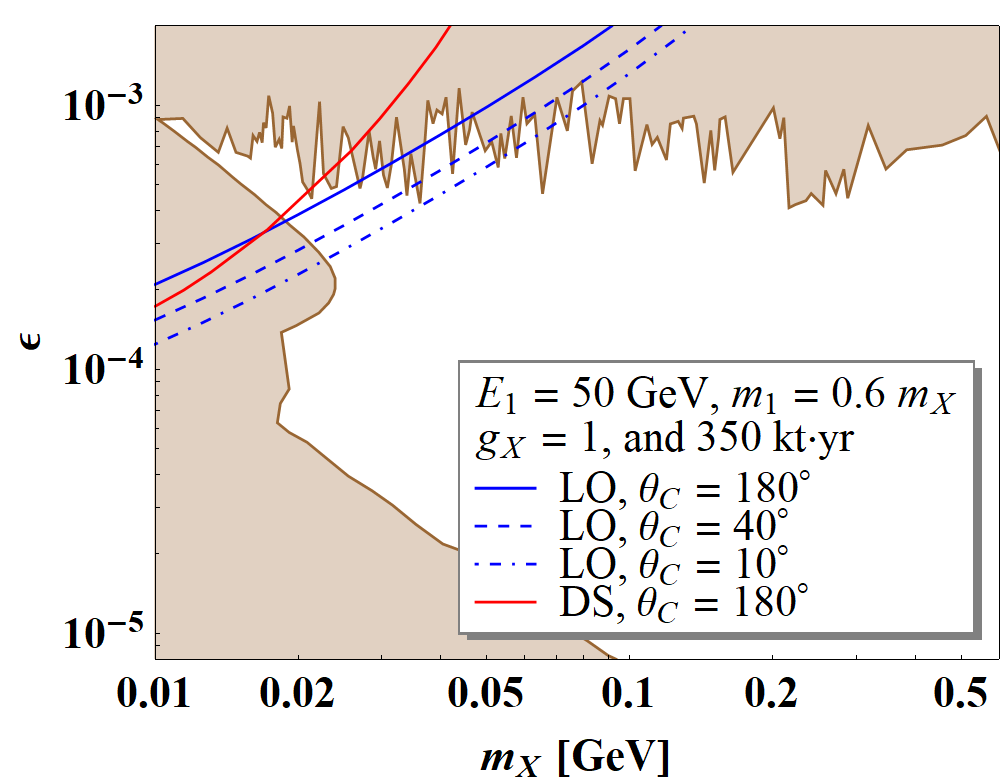}

\vspace{0.5cm}

\includegraphics[width=8.0cm]{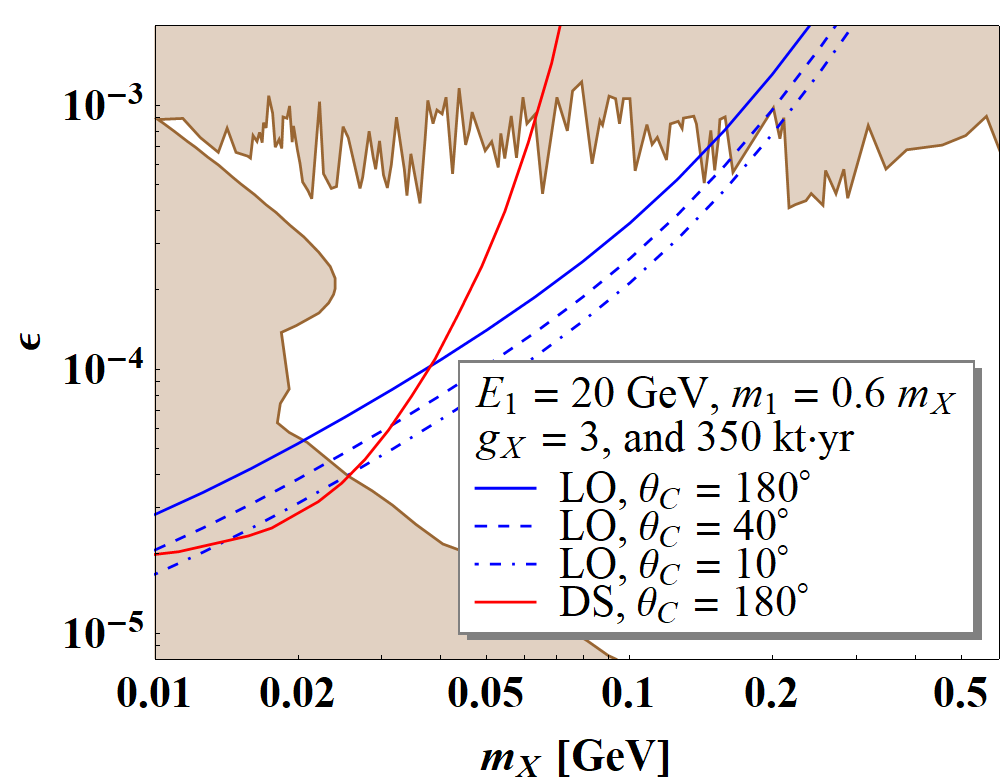}\hspace{0.5cm}
\includegraphics[width=8.0cm]{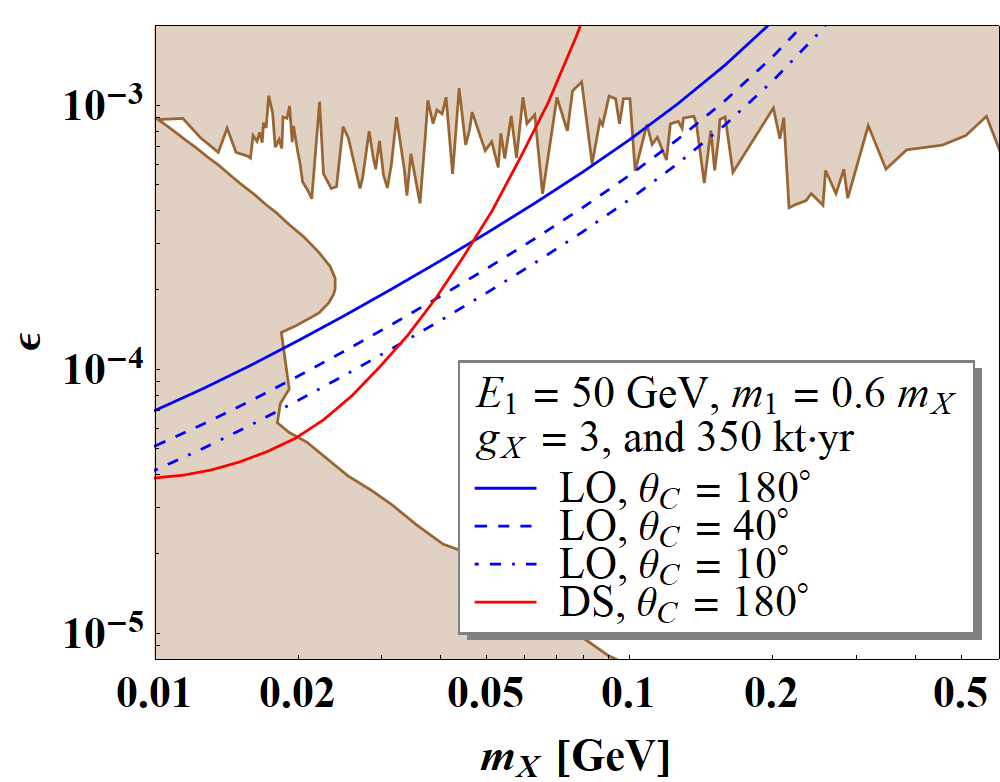}
\caption{\label{fig:sensitivites_DUNE}
Experimental sensitivities of DUNE in the dark gauge boson model parameters $m_X - \epsilon$ for the cases of leading-order (blue) and dark-strahlung-induced (red) scattering processes, of 350 kt$\cdot$yr exposures. 
The top (bottom) panels are for $g_X=1$ ($g_X=3$) while the left (right) panels are for $E_1=20$ GeV ($E_1=50$ GeV).
A mass hierarchy of $m_1=0.6\, m_X$ is kept for all results.
The brown-shaded regions show the currently excluded parameter space according to the report in Ref.~\cite{Banerjee:2018vgk}.
For the dark-strahlung channel, we assume the all-sky survey (i.e., $\theta_C=180^\circ$) and negligible background, whereas for the leading-order, we consider two more search cones, $\theta_C=10^\circ$ (dot-dashed lines) and $\theta_C=40^\circ$ (dashed lines), in addition to the all-sky survey (solid lines).
}
\end{figure*}

We are now in the position to investigate how the main virtue of dark-strahlung is realized under the realistic experimental circumstances.
The primary purpose of the discussion here is to develop our intuition on the impact of realistic effects, so we do not pursue any dedicated detector-level studies.
Since we adopted a dark gauge boson scenario, it is natural enough to study and compare experimental sensitivities through the leading-order and dark-strahlung channels in the standard $m_X-\epsilon$ plane.
Again, events are generated by \texttt{MG5\_aMC@NLO}, and the subsequent analysis is conducted under the DUNE far-detectors setup, as mentioned earlier.
Therefore, we require electrons to exceed the associated threshold, i.e., $E_{e^\pm}>30$ MeV~\cite{Acciarri:2015uup}, and the recoil electron to be separated from the dark-strahlung-induced $e^{\pm}$ by at least $1^\circ$~\cite{Acciarri:2015uup}.
We additionally demand the primary boosted $\chi_1$ scattering and the secondary decay of the emitted $X$ to take place inside the detector fiducial volume.
Indeed, we find that $X$ can be {\it appreciably displaced}, given the position resolution of $\sim 1$ cm~\cite{Acciarri:2015uup} at the DUNE detectors.
More quantitatively, the laboratory-frame mean decay length is given by~\cite{Giudice:2017zke}
\bea
\ell_{X,{\rm lab}}\sim 2\hbox{ cm}\times \left(\frac{10^{-4}}{\epsilon}\right)^2 \left( \frac{40 \hbox{ MeV}}{m_X}\right)\frac{\gamma_X}{100}\,.
\eea
To calculate the acceptances in $\ell_{X,{\rm lab}}$, we follow the conservative scheme suggested in Ref.~\cite{Giudice:2017zke}, taking the dimension of the fiducial volume for a single module of the DUNE far-detectors.

FIG.~\ref{fig:sensitivites_DUNE} displays the expected 90\% C.L. sensitivities of DUNE in the dark gauge boson model parameters $m_X - \epsilon$ for the cases of leading-order (blue) and dark-strahlung-induced (red) scattering processes, of 350 kt$\cdot$yr exposures. 
Taking the entire galatic halo as the origin of the $\chi_1$ flux, we closely follow the analysis procedure detailed in Ref.~\cite{Giudice:2017zke}.
The legends contain our parameter choices, and the up-to-date exclusion limits (brown-shaded regions) are from Ref.~\cite{Banerjee:2018vgk}.
For the dark-strahlung channel, we assume the all-sky survey (i.e., $\theta_C=180^\circ$) and negligible background, whereas for the leading-order, we consider two more search cones, $\theta_C=10^\circ$ (dot-dashed lines) and $\theta_C=40^\circ$ (dashed lines) as per suggestions in Refs.~\cite{Kachulis:2017nci, Agashe:2014yua}, in addition to the all-sky survey (solid lines).
The number of neutrino-induced background events for the leading-order process is calculated with Eqs.~\eqref{eq:Nallsky} and~\eqref{eq:Ntheta}.
Our results suggest that the dark-strahlung channel is at least {\it complementary} to the leading-order ones.
Indeed, dark-strahlung allows us to explore a wider range of parameter space, especially the parameter regions towards small $m_X$, than even the leading-order channels with a search cone.
We also clearly observe that the curvatures for the dark-strahlung in the bottom panels change a lot towards smaller $\epsilon$ and/or smaller $m_X$, since the vertex displacement becomes comparable to the detector characteristic length scale, hence the acceptance is rather degraded.

Finally, be aware that our event selection scheme does not require any angular separation between the dark-strahlung-induced $e^\pm$'s.
Indeed, the $e^+e^-$ pair is often merged, so resolving them would necessitate more dedicated techniques, e.g., $dE/dx$-analysis~\cite{Acciarri:2016sli,Acciarri:2018ahy} and machine-learning-based algorithm~\cite{Acciarri:2016ryt,Adams:2018bvi}, which is beyond the scope of our study.
Moreover, as discussed above, many dark-strahlung-induced events are expected to show the delayed decay of $X$ so that the identification of a time-correlated displaced vertex would suffice to tag them.

\section{Conclusions and Outlook \label{sec:conclusion}}
In this study, we investigated the novel search channel for the BDM signals coming from the present universe, utilizing {\it dark-strahlung}, a higher-order contribution, by which boosted dark matter radiates massive dark gauge bosons.
The resultant final state consists of a visible target recoil along with decay products of the emitted dark gauge boson.
The uniqueness of the experimental signature renders associated searches essentially background-free.
Due to its higher-order nature, however, the dark-strahlung-associated channel [consisting of both FSDS and ISDS in FIG.~\ref{fig:diagram}($b$)] may be in tension with relatively smaller signal statistics than the leading contribution [i.e., the first diagram in FIG.~\ref{fig:diagram}($b$)].

In light of this situation, we showed that if BDM comes with a significant boost factor, production rates of the dark-strahlung-induced events can be substantially enhanced.
Moreover, if backgrounds to the leading-order process are large, the dark-strahlung channel can play a prominent role.
We argued that the cosmic-origin BDM searches fall in this category, and therefore they can benefit enormously from dark-strahlung.
We further studied the impact of realistic experimental environment on the usefulness of dark-strahlung, using DUNE far-detectors as an example.
The capabilities of both leading-order and dark-strahlung channels for probing dark gauge boson parameter space were contrasted.
Remarkably enough, the dark-strahlung channel may allow for more stringent limits in a wide range of parameter space.

In conclusion, the channel with dark-strahlung can be {\it complementary} to, or even surpass the corresponding leading-order one, especially in cosmogenic signal searches.
Furthermore, its existence can be important evidence to refute the candidate hypothesis that the signals are induced by neutrinos originating from the decay/pair-annihilation of halo dark-matter.

Beyond our study, signatures induced by a multitude of dark-strahlung will provide richer phenomenology.
Although some experimental challenges to signal reconstruction are expected~\cite{deGouvea:2018cfv}, it is certainly an interesting research direction to pursue.
Finally, although our study was performed solely with electrons for illustration, one is readily embarked on similar examinations for the proton scattering channels.
We reserve more dedicated analyses in this direction for a forthcoming work~\cite{futurework}.

\section*{Acknowledgments}
We thank Lucien Heurtier and Olivier Mattelaer for useful discussions.
DK is supported by the Department of Energy under Grant No. DE-FG02-13ER41976/DE-SC0009913.
JCP is supported by the National Research Foundation of Korea (NRF-2016R1C1B2015225 and NRF-2018R1A4A1025334).
SS is supported by the National Research Foundation of Korea (NRF-2017R1D1A1B03032076).

\end{document}